\documentclass[pdflatex,sn-mathphys-num]{sn-jnl}% Math and Physical Sciences Numbered Reference Style
\usepackage{bm}
\usepackage{bbold}
\usepackage{graphicx}
\usepackage{svg}
\usepackage{color}
\usepackage[binary,amssymb]{SIunits}
\usepackage{comment}
\usepackage{hyperref}
\usepackage{amsmath,amssymb}
\usepackage[capitalise]{cleveref}
\usepackage{color}

\newcommand{\si}[1]{\mathrm{~{#1}}}

\begin{document}

\title[Emergent Quantum Geometric Phases in Holey Graphene]
{Emergent Quantum Geometric Phases in Holey Graphene}

\author[1,2]{\fnm{Pablo} \sur{Canteli}}\email{p.canteli@usal.es}
\author[3]{\fnm{Yuriko} \sur{Baba}}\email{yuriko.baba@csic.es}
\author[1,4]{\fnm{Juan} \sur{Salvador-S\'{a}nchez}}\email{juan2s@usal.es}
\author[5]{\fnm{Jorge} \sur{Estrada-\'{A}lvarez}}\email{jorgestr@ucm.es}
\author[1,2]{\fnm{Ana} \sur{Pérez-Rodríguez}}\email{perez.rodriguez.ana@usal.es}
\author[1]{\fnm{Carlos} \sur{S\'{a}nchez-S\'{a}nchez}}\email{carlos.sanchez@usal.es}
\author[1]{\fnm{Vito} \sur{Cleric\`{o}}}\email{vito$\_$clerico@usal.es}
\author[6]{\fnm{Takashi} \sur{Taniguchi}}
\author[7]{\fnm{Kenji} \sur{Watanabe}}
\author[5]{\fnm{Francisco} \sur{Domínguez-Adame}}\email{adame@ucm.es}
\author[3]{\fnm{Rafael A} \sur{Molina}}\email{rafael.molina@csic.es}
\author[1]{\fnm{Enrique} \sur{Diez}}\email{enrisa@usal.es}
\author[5]{\fnm{Elena} \sur{D\'{i}az}}\email{elenadg@fis.ucm.es}
\author*[1,2]{\fnm{Mario} \sur{Amado}}\email{mario.amado@usal.es}

\affil*[1]{\orgdiv{Nanotechnology Group, USAL-Nanolab}, Departamento de Física Fundamental, \orgname{Universidad de Salamanca}, \orgaddress{ \city{Salamanca}, \postcode{E-37008}, \country{Spain}}}
\affil*[2]{\orgdiv{IUFFyM}, \orgname{Universidad de Salamanca}, \orgaddress{ \city{Salamanca}, \postcode{E-37008}, \country{Spain}}}
\affil[3]{\orgname{Instituto de Estructura de la Materia IEM-CSIC}, \orgaddress{ \city{Madrid}, \postcode{E-28006}, \country{Spain}}}
\affil[4]{\orgdiv{Unidad de Excelencia en Luz y Materia Estructurada (LUMES)}, \orgname{Universidad de Salamanca}, \orgaddress{ \city{Salamanca}, \postcode{E-37008}, \country{Spain}}}
\affil[5]{\orgdiv{Departamento de Física de Materiales}, \orgname{Universidad Complutense}, \orgaddress{ \city{Madrid}, \postcode{E-28040}, \country{Spain}}}
\affil[6]{\orgdiv{Research Center for Materials Nanoarchitectonics}, \orgname{ National Institute for Materials Science}, \orgaddress{ \city{Tsukuba}, \postcode{305-0044}, \country{Japan}}}
\affil[7]{\orgdiv{Research Center for Electronic and Optical Materials}, \orgname{ National Institute for Materials Science}, \orgaddress{ \city{Tsukuba}, \postcode{305-0044}, \country{Japan}}}

\abstract{

In graphene and other two-dimensional materials, periodic 
modulations of the electron density can significantly alter 
the energy spectrum and transport properties. Here, we report magnetotransport measurements in encapsulated monolayer graphene with ultra–high-quality patterned periodic antidot lattices that preserve the intrinsic electronic properties of the material. This lithographically defined platform enables controlled access to commensurability and superlattice phenomena at length scales otherwise difficult to achieve. By systematically tuning the lattice dimensions, we reveal a hierarchy of classical commensurability features arising from cyclotron orbits with comparable radii that follow multiple classical trajectories, resulting in broadened resistance peaks beyond the conventional single-orbit picture. Superimposed on these features, we observe pronounced Brown-Zak oscillations arising from the quantum commensurability between the magnetic flux quantum and the unit cell of the engineered Bravais lattices. We demonstrate that the intrinsic geometric phase of our system is directly measurable and show a precise matching of the magnetic field periodicity to the lithographic periodic patterning, where moiré-like electronic spectra can be geometrically generated in single-layer graphene without the need for twist, lattice mismatch, or multilayer stacking. Our results establish nanopatterned graphene as a clean, tunable, and scalable platform for realizing and exploring moiré physics through on-demand real-space design.
}

\keywords{single layer graphene, moiré superlattice effect, antidot lattice, geometric patterning}

\maketitle

\section{Introduction}\label{Introduction}

Periodic modulations in two-dimensional (2D) materials such as graphene profoundly reshape their electronic spectrum, leading to the formation of minibands~\cite{Wallbank2013,Mrenca2023}, commensurability oscillations~\cite{Yagi2015,Sandner2015,Power2017,Drienovsky2018}, and fractal energy spectra such as the Hofstadter butterfly in the presence of a magnetic field~\cite{Nemec2007}. In graphene, these effects arise when a periodic potential folds the Dirac cone and reconstructs the low-energy band structure. Such modulations have been realized using a variety of strategies, including chemical patterning~\cite{Sun2011}, self-assembled nanostructures~\cite{Zhang2018}, and naturally occurring moiré superlattices in aligned or slightly twisted graphene/hexagonal BN (\textit{h}-BN) heterostructures~\cite{Xue2011,Du2020,Salvador2024,Ponomarenko2013,Dean2013,Wang2025,Juan2024}. These moiré superlattices in 2D-materials generate long-wavelength periodic potentials that have enabled the observation of the Hofstadter butterfly~\cite{Dean2013,Ponomarenko2013}, Brown--Zak oscillations~\cite{Kumar2017,Barrier2020,Huber2022}, and a wealth of correlated phenomena in multilayer systems, including magic-angle graphene~\cite{Cao2018}.\\

While moiré superlattices are typically associated with lattice mismatch or relative twist in van der Waals heterostructures~\cite{Yankowitz2012,Ponomarenko2013,Dean2013,Cao2018,Wang2019,Ezzi2024}, they can also be generated artificially via periodic real-space nanopatterning. In particular, antidot lattices impose a spatially periodic modulation that acts as an effective moiré potential~\cite{Weiss1991,Weiss1993}, thereby bypassing the requirement for structural crystallographic alignment~\cite{Mrenca2023,Barcons2022}. When combined with external electrostatic gating, such nanopatterned systems enable the creation of synthetic 2D band structures whose electronic properties can be tuned independently of the host lattice~\cite{Wang2023}. Beyond simple carrier confinement, these engineered potentials introduce a distinct interplay between classical and quantum magnetotransport. On one hand, they manifest as commensurability features, Weiss oscillations, of predominantly classical origin, dictated by the geometric matching between ballistic cyclotron orbits and the superlattice periodicity~\cite{Weiss1991,Weiss1993,Ishizaka1997,Yagi2015,Sandner2015,Power2017,Drienovsky2018}. The latter have been found to be extraordinarily robust even in disordered samples due to an intrinsic linking between collision times and accessible phase space volumes following Kac's lemma~\cite{Datseris2019}. On the other hand, the periodic nanopatterned potential restores discrete translational symmetry in a magnetic field, giving rise to the abovementioned quantum-mechanical Brown--Zak oscillations~\cite{Brown1964,Zak1964a,Zak1964b} that were first observed in GaAS heterostructures~\cite{Weiss1991,Weiss1993} and lately reported in graphene-heterostructures~\cite{Kumar2017,Barrier2020,Huber2022}. This phenomenon is fundamentally governed by the quantization of the magnetic flux per unit cell area, occurring when the flux matches rational fractions of the magnetic flux quantum $\phi_0 = h/e$, where $h$ is Planck’s constant and $e$ is the elementary charge. In graphene, the coexistence of high mobility, electrostatic tunability, and engineered periodic potentials therefore provides a powerful platform to explore the boundary between classical cyclotron dynamics and quantum-coherent miniband transport.\\

In this work, we demonstrate that controlled nanopatterning alone provides a robust pathway to engineer moiré-like superlattices in graphene, where classical and quantum transport phenomena coexist. We report magnetotransport measurements in high-mobility \textit{h}-BN encapsulated graphene featuring periodic arrays of antidots with diameters ($\Theta$) ranging from $50$ to $300\,$nm. This periodic potential is defined deterministically using electron-beam lithography (EBL), enabling precise control over the superlattice geometry and energy scales without invoking twist angles or intrinsic lattice mismatch. We distinctively identify both classical Weiss-type commensurability peaks and quantum Brown--Zak oscillations, demonstrating that our lithographic unit cell is fully capable of resolving the magnetic flux quantization per unit area. Our results establish nanopatterned graphene as a versatile platform for moiré engineering, highlighting lithographic patterning as a general and scalable route to design superlattices beyond naturally occurring moiré systems.

\section{Results}\label{Results}

\begin{figure*}[htbp]
\centering
\includegraphics[width = 1\linewidth]{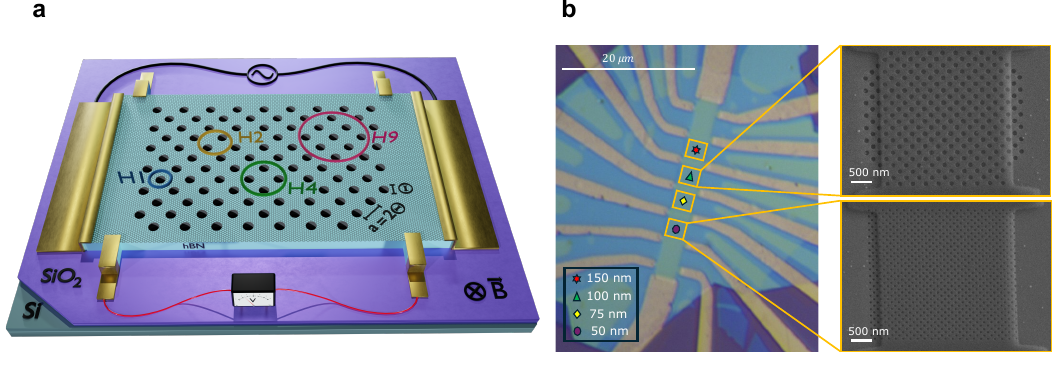}
\caption{\textbf{Device schematic and optical micrograph for Sample A.} \textbf{a}) Schematic illustration of a simplified Hall bar geometry showing the source, drain, and voltage contacts. The antidots form a rotated square lattice, where the colored circles depict classical cyclotron orbits encircling 1, 2, 4, and 9 antidots, representing the primary non-chaotic contributions of Weiss oscillations to the longitudinal resistance. \textbf{b}) Optical micrograph of a 16-terminal Hall bar device. The device features four distinct patterned regions with antidot diameters of $\Theta = 50,\,75,\,100,$ and $150\,\rm nm$. Two unpatterned, pristine regions at the ends of the Hall bar serve as reference sections to characterize the unperturbed electrical response. The insets show scanning electron micrographs of fully etched antidot arrays ($\Theta = 50$ and $100\,\rm nm$) performed on sacrificial \textit{h}-BN flakes.}\label{superfigure1}
\end{figure*}

\subsection{Device design and magnetotransport characterization}
Fully encapsulated monolayer graphene heterostructures were assembled on a \textit{p}-doped silicon wafer featuring a $300\,$nm top layer of silicon oxide. The constituent flakes were prepared via standard mechanical exfoliation of pristine graphite and \textit{h}-BN crystals (see Methods for more information). To pattern the antidot arrays, a cryo-etching technique was employed to minimize edge roughness~\cite{Clerico2019,EBLVito2020}. This ensures clean ballistic transport by etching through the top \textit{h}-BN and underlying graphene layer while leaving the silicon backgate intact. Two distinct devices, hereafter referred to as Sample~A and Sample~B, were fabricated with the antidots arranged in a square lattice rotated by 45$^\circ$ relative to the transport channel. Across both samples, the closest-neighbor center-to-center lattice spacing $a$ is fixed at twice the diameter of the antidot, i.e. $a = 2\Theta$. \\

The completed devices thus integrate segments of pristine graphene alongside distinct patterned antidot superlattices of varying dimensions. As illustrated in the schematic of Fig.~\ref{superfigure1}(a), these superlattices are positioned directly between the longitudinal voltage probes. Specifically, Sample A features regions with diameters of $\Theta = 50\,, 75\,, 100,$ and $150\,\rm nm$, whereas Sample B incorporates larger diameters of $\Theta = 100\,, 200$ and $300\,\rm nm$. The Hall bar channels maintain a uniform width $W = 3\,\mu\rm m$ and a longitudinal voltage probe spacing of $L = 3\,\mu\rm m$, yielding a constant aspect ratio of $W/L = 1$ across all active regions. Initial electronic characterization demonstrated excellent device quality, with both samples exhibiting a high carrier mobility of $\mu \sim 10^5\,\rm cm^2/Vs$ at the pristine regions and a low residual charge carrier density of $n_c^* \sim 8\times10^{10}\,\rm cm^{-2}$. The mean free path was estimated using $l_{\rm{mfp}}= \mu\sqrt{\pi\vert n\vert}\hbar/e$  yielding $l_{\rm{mfp}} \gtrsim 1\,\mu$m in the pristine zone for the carrier densities relevant to our devices, ensuring that undesired impurity or phonon scattering does not dominate the transport properties~\cite{Rossi2011}. The antidot density scale set by the patterned periodic potential is more than an order of magnitude smaller than the residual carrier density. Consequently, the superlattice-induced satellite Dirac points cannot be resolved and remain hidden beneath the primary Dirac peak.
\\

\subsection{Commensurability effect}
\begin{figure}[h!]
\centering
\includegraphics[width = 1\linewidth]{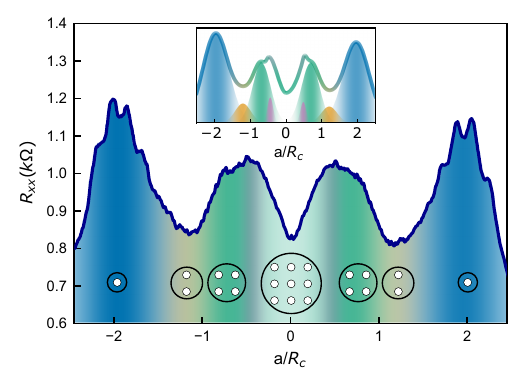}
\caption{\textbf{Magnetoresistance measurement and commensurability peak identification.} Longitudinal magnetoresistance $R_{xx}$ measured as a function of the normalized magnetic field $a/R_C$ (where $R_C$ is the cyclotron radius) for the $\Theta=75\,\rm nm$ region of Sample A. The data were acquired at $T=1.5\,\rm K$ under a fixed carrier density of $n_c=1.81\times10^{12}\,\rm cm^{-2}$. Distinct commensurability features are resolved, corresponding to classical cyclotron orbits that encircle 1, 2, 4, and 9 antidots, with their respective radii determined via Eq.~\eqref{eq:commensurability}. We will label these orbits by $\mathrm{H} k$, being $k$ the number of enclosed antidots, i.e. $\rm H1$, $ \rm H2$, $\rm H4$ and $\rm H9$. The outermost and most prominent commensurability peak $\rm H1$ emerges at $a \approx 2R_C$, consistent with a classical trajectory pinned around a single antidot. Inset: Lorentzian fits isolating the individual contribution of each commensurability peak to the overall longitudinal resistance.}
\label{normalizedcomm}
\end{figure}

Antidot superlattices provide a robust macroscopic platform for investigating semiclassical cyclotron dynamics. When subjected to a perpendicular magnetic field, charge carriers in these artificial periodic potentials map out characteristic classical trajectories that manifest as distinct commensurability (Weiss) peaks in the longitudinal magnetoresistance. As shown in Fig.~\ref{normalizedcomm}, the exceptional electronic quality of our patterned encapsulated graphene devices enables the resolution of multiple, distinct families of these peaks~\cite{Ishizaka1997,Power2017}. We classify these orbit families as $\mathrm{H}k$, where $k$ denotes the number of antidots enclosed by the semiclassical cyclotron trajectory. Specifically, we resolve higher-order geometric paths corresponding to $\rm H1$, $\rm H2$, $\rm H4$, and $\rm H9$ configurations across various carrier densities and patterned regions. \\

The magnetic field $B_C$ at which a given commensurability condition is satisfied is governed by the carrier density $n_c$ and the classical cyclotron radius $R_C$ through the fundamental relation~\cite{Drienovsky2018}:
\begin{equation}\label{eq:commensurability}
B_\mathrm{C} = \frac{\hbar}{eR_\mathrm{C}} \sqrt{\pi |n_c|}\ .
\end{equation}
The main resonance, $\mathrm{H1}$, satisfies the geometric boundary condition $a \approx 2R_\mathrm{C}$, corresponding to a trajectory securely pinned around a single antidot~\cite{Power2017}. While this primary condition is independent of the lattice geometry, the higher-order spatial configurations, where trajectories enclose multiple scatterers or form skipping orbits depend strictly on the square lattice arrangement. Because the scaling ratio $a = 2\Theta$ is preserved in all devices, these geometric commensurability conditions remain consistent in all patterned regions.\\

Macroscopically, these transport features originate from the fundamental coexistence of chaotic dynamics and regular pinned trajectories in systems governed by Kac's recurrence theorem~\cite{Kac1947}. In the framework of a magnetic Sinai billiard, Kac's theorem enforces the conservation of total phase-space volume, causing chaotic and regular trajectories to continuously compete for accessible phase space~\cite{Weiss1991,Datseris2019,Ponomarenko2008}. When the external magnetic field tunes the cyclotron radius to match a stable geometric condition, pinned periodic orbits form. These stable orbits effectively exclude chaotic trajectories from their phase-space volume. Consequently, the mean collision time for carriers within the chaotic sea decreases. Following the Drude model, this enhanced effective scattering translates directly into an increased longitudinal resistance, producing the characteristic commensurability maxima that we are able to observe.\\

To disentangle the contributions of the different $\mathrm{H}k$ families, the magnetoresistance traces can be phenomenologically decomposed into a sum of Lorentzian functions centered at their respective geometric commensurability fields (Fig.~\ref{normalizedcomm}, inset). Because an exact analytical expression for the Weiss peak lineshape remains unknown, the Lorentzian profile serves as an empirical approximation. Crucially, the finite width of these fits reflects the underlying existence of quasi-stable trapped trajectories. These orbits repeatedly scatter between adjacent antidots, maintaining a temporary confinement that encloses the same effective number of scatterers before eventually escaping into the chaotic sea~\cite{Datseris2019}. This transient trapping explains why the commensurability tails extend into nominally incommensurate magnetic-field regimes. Finally, it is essential to note that the visibility of a pronounced commensurability peak is not a strict prerequisite for the existence of its corresponding pinned trajectory family. Depending on the carrier density, the contribution of a specific higher-order commensurability family may become too weak to produce an isolated resistance maximum. \\

In our measurements, on top of the Weiss commensurability peaks, distinct fine-structure wrinkles (that will be identified as Brown-Zak oscillations) consistently emerge within the magnetic-field intervals associated with these $\mathrm{H}k$ conditions. Moreover, we show that these underlying classical trajectories act as dynamical visibility windows, strictly governing the quantum transport properties even when their individual semiclassical peaks are not independently resolved.\\

The dependence of the longitudinal resistance $R_{xx}$ on the magnetic field for different carrier densities, antidots diameters and temperature is reported in Figure~\ref{classic}. Unless otherwise stated, data at $T=1.5$K will be shown. 
Panel (a) displays a contour plot of $R_{xx}$ as a function of the external magnetic field and carrier density, measured for the $\Theta=75\,\rm nm$ region of Sample A. 
The gray-shaded area near the charge neutrality (Dirac) point is omitted to optimize the contrast of the color scale. The center of the primary commensurability H1 peak is tracked across densities by the dashed line following Eq.~\eqref{eq:commensurability}. At moderate magnetic fields exceeding the values required to reach the primary commensurability peak, clear Shubnikov-de Haas (SdH) oscillations become notably visible. These oscillations follow the expected linear relationship originating from the $(n_c,B)=(0,0)$ intercept, characteristic of high-field quantum transport.\\ 

Panel (b) compares the magnetoresistance profiles for various antidot dimensions across Samples A and B as a function of the normalized magnetic field scale $a/R_C$. As expected, all traces exhibit a highly consistent scaling behavior, marked by a prominent coincident H1 peak at $a=2R_C$ and demonstrates excellent consistency between all regions of the two fabricated samples. \\
 
Panel (c) presents $R_{xx}$ as a function of the magnetic field $B$ for the $\Theta=100\,\rm nm$ region of Sample B over a carrier density range of $n_c=0.50 \times 10^{12}\,\rm cm^{-2}$ to $2.25 \times 10^{12}\,\rm cm^{-2}$ (data for other antidot dimensions can be found in Extended Data). Distinct fine-structure oscillations, or \emph{wrinkles}, are superimposed across all commensurability peaks and mostly visible on H1, as seen in Fig.~\ref{normalizedcomm}. Finally, Fig.~\ref{classic}(d) illustrates the temperature evolution of the longitudinal resistance of the $\Theta=100\,\rm nm$ superlattice region of Sample A at a fixed carrier density of $n_c = 0.45 \times 10^{12}\,\rm cm^{-2}$. While the primary commensurability peak remains robust as temperature increases, these fine-structure wrinkles are progressively suppressed by thermal smearing and are entirely washed out for $T \gtrsim 15\,\rm K$.

\begin{figure*}[htbp]
\centering
\includegraphics[width = 1\linewidth]{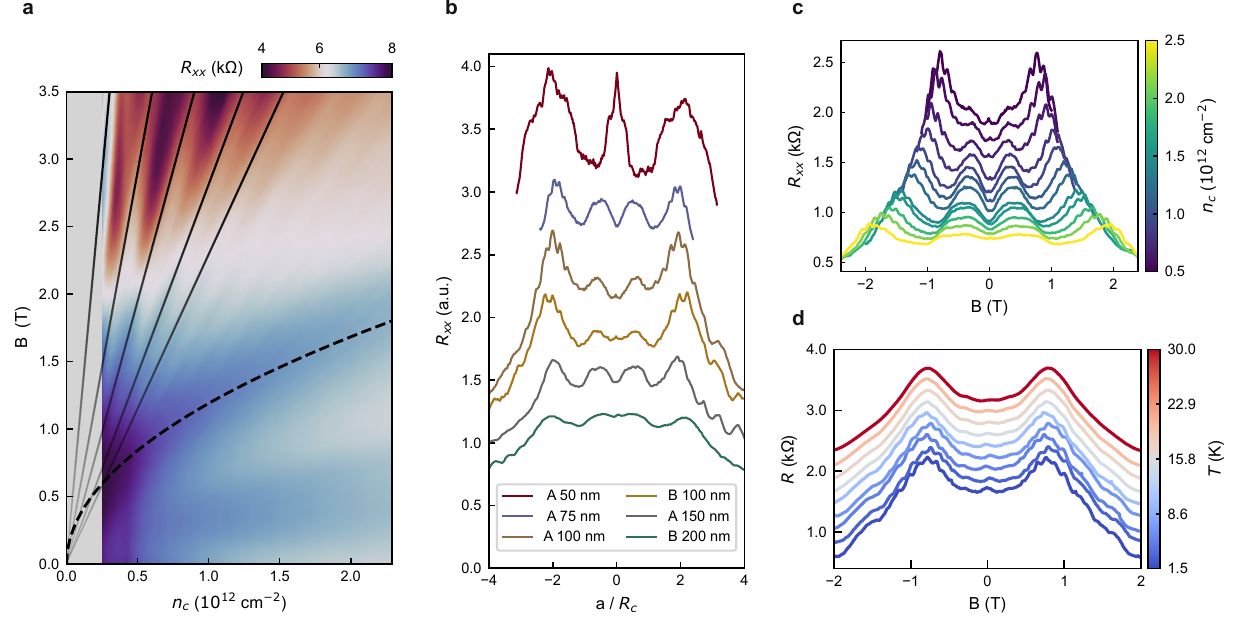}

\caption{\textbf{Density dependence, scaling behavior, and thermal suppression of commensurability features.} \textbf{a)}~Contour plot of the longitudinal resistance $R_{xx}$ as a function of external out-of-plane magnetic field $B$ and carrier density $n_c$ for the $\Theta = 75\,\rm nm$ antidot region of Sample A. The gray-shaded area near the Dirac point is omitted to enhance contrast. The dashed line tracks the evolution of the primary commensurability peak, %while the dominant linear branches at higher fields correspond to standard Shubnikov-de Haas oscillations. 
while the linear branches corresponding to standard Shubnikov-de Haas oscillations become notably visible at higher fields.
\textbf{b)}~$R_{xx}$ as a function of the normalized magnetic field $a/R_C$ for different antidot diameters across regions of both Sample A and Sample B. The alignment of the prominent resistance peak at $a = 2R_C$ demonstrates excellent geometric scaling across all devices. \textbf{c)}~$R_{xx}$ sweeps for the $\Theta = 100\,\rm nm$ region of Sample B at various carrier densities ranging from $n_c=0.50 \times 10^{12}\,\rm cm^{-2}$ to $2.25 \times 10^{12}\,\rm cm^{-2}$, showing fine-structure \emph{wrinkles} mostly visible superimposed on the main commensurability peak. All data obtained in panfels \textbf{a} to \textbf{c} were acquired at $T=1.5\,\rm K$. \textbf{d)}~Temperature-evolution $R_{xx}$ profiles for the $\Theta=100\,$nm region of Sample A at a fixed carrier density of $n_c = 0.45 \times 10^{12}\,\rm cm^{-2}$ (curves are vertically offset by $200\,\Omega$ for clarity), illustrating the progressive thermal smearing of the fine-structure ripples.}\label{classic}\label{classic}
\end{figure*}

Notice that the longitudinal magnetoresistance within the region patterned with the smallest antidots ($\Theta = 50\,\rm nm$) exhibits a pronounced weak localization peak around zero magnetic field~\cite{Tikhonenko2008,Pezzini2012}. This localized feature is driven by enhanced intervalley scattering occurring at the etched boundaries of the structures, which accounts for the strong localization signatures typically observed in superlattices with short spatial periods~\cite{Eroms2009}. Crucially, this region also resolves clearly the outermost commensurability feature $\rm H1$, demonstrating that even at these reduced dimensions, the periodic geometric modulation continues to exert a dominant influence on the magnetotransport response.\\

In conventional antidot arrays, commensurability peaks are expected to suppress rapidly at low carrier densities due to the breakdown of classical transport as the Fermi wavelength approaches the spatial constriction width. Specifically, Sandner \emph{et al.}~\cite{Sandner2015} proposed that maintaining classical trajectories requires satisfying the strict condition $\lambda_F / 2\pi \ll D$, where $\lambda_F = 2\sqrt{\pi/n_c}$ is the Fermi wavelength and $D$ is the edge-to-edge constriction width ($D = \Theta = 50\,\rm nm$ in our smallest geometry). Contrary to this criterion, our devices resolve distinct Weiss peaks deep into the low-density regime, down to $n_c = 0.45 \times 10^{12}\,\rm cm^{-2}$, where this inequality clearly no longer holds. This persistence demonstrates that the classical-to-quantum transition is far more gradual than previously assumed, revealing an unexpected robustness of semiclassical trajectories at long Fermi wavelengths.\\

This observation also addresses the key transport bottleneck identified by Jessen \emph{et al.}~\cite{Jessen2019}, who argued that aggressive feature downscaling typically introduces edge roughness that degrades mobility and drives the system away from purely ballistic transport. In our heterostructures, despite scaling down to a minimum feature size of $\Theta = 50\,\rm nm$, the structural cleanliness of the patterned regions suppresses diffusive edge scattering and keeps carriers firmly within the ballistic limit, providing a clear physical rationale for why robust commensurability peaks persist in our devices where earlier architectures struggled.

%%%%%%%%%%%%%%%%%%%%%%%%%%%%%%%%%%%%%%%%%%%%%
\subsection{Hierarchical Self-Similarity and Geometric Flux Quantization}
%%%%%%%%%%%%%%%%%%%%%%%%%%%%%%%%%%%%%%%%%%%%%

\begin{figure*}[htbp]
    \centering
    \includegraphics[width=1\textwidth]{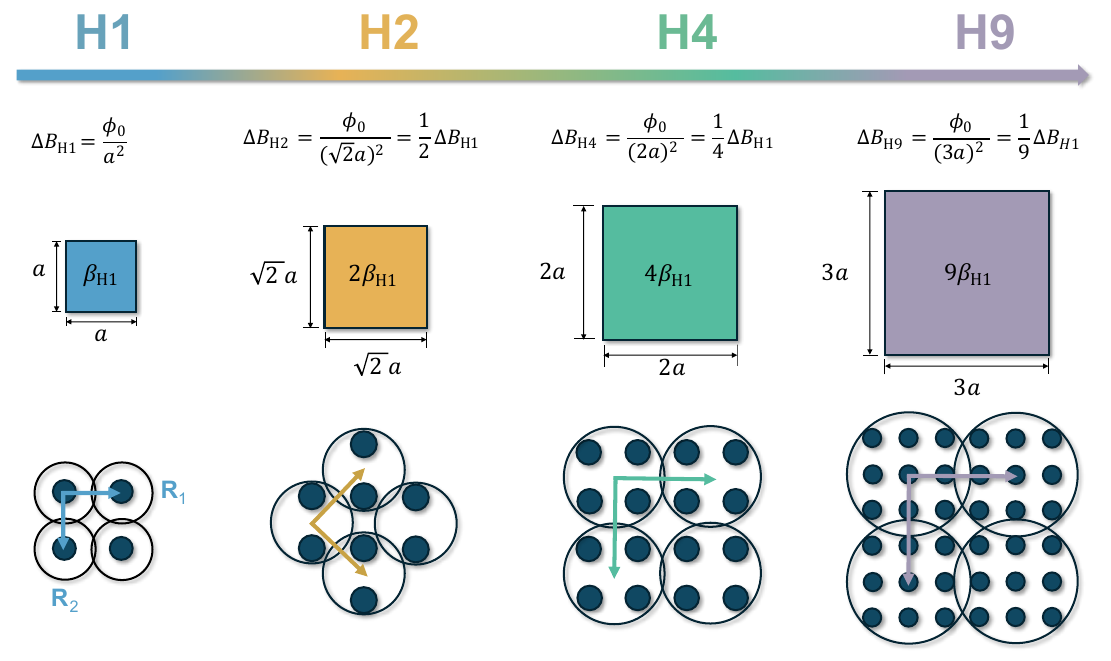}
   \caption{\textbf{Hierarchical self-similarity and flux-quantization rescaling in the square antidot superlattice.} Schematic representation of the nested, commensurate sub-lattices ($\mathrm{H1}$, $\mathrm{H2}$, $\mathrm{H4}$, and $\mathrm{H9}$) formed by the collective behavior of carrier trajectories enclosing multiple scatterers. \textbf{H1}~The fundamental square unit cell of side length $a$ defining the baseline field spacing required to insert one extra flux quantum, $\Delta B_{\mathrm{H1}} = h/(ea^2) = \phi_0 n_1$. This establishes the fundamental flux ratio $\beta_{\mathrm{H1}} = \phi/\phi_0$ for the primary unit cell. \textbf{H2}~The first-order expanded sub-lattice, rotated by $45^{\circ}$ with an effective period of $\sqrt{2}a$. The real-space unit-cell area doubles ($S_2 = 2a^2$), effectively halving the antidot density probed by the carriers ($n_2 = n_1/2$), doubling the enclosed flux ratio to $\beta_{\mathrm{H2}} = 2\beta_{\mathrm{H1}}$, and halving the characteristic magnetic field spacing to $\frac{1}{2}\Delta B_{\mathrm{H1}}$. \textbf{H4, H9}~Higher-order geometric expansions enclosing $2a \times 2a$ and $3a \times 3a$ superlattice domains, respectively. The quadratic increase in the enclosed unit-cell area ($S_k = k a^2$) yields a discrete, fractional rescaling of the effective density $n_k = n_1/k$, scales the flux ratio to $\beta_{\mathrm{H}k} = k\beta_{\mathrm{H1}}$, and shifts the flux-quantization interval to $\frac{1}{4}\Delta B_{\mathrm{H1}}$ and $\frac{1}{9}\Delta B_{\mathrm{H1}}$.}
    \label{fig:Zak}
\end{figure*}

The prominent, equidistant fine-structure oscillations superimposed on the commensurability (Weiss) peaks~\cite{Eroms2009,Sandner2015} are attributed to Brown-Zak oscillations. These features, becoming particularly pronounced on top of the outermost Weiss oscillation ($\rm H1$), cannot be attributed to either Shubnikov-de Haas oscillations or classical unpinned cyclotron trajectories; we interpret them as a direct signature of the strong, spatially periodic potential induced by the antidot superlattice.
While similar features have been observed in conventional 2D electron gases~\cite{Weiss1991,Weiss1993} and graphene heterostructures~\cite{Kumar2017,Barrier2020,Huber2022}, they have historically lacked comprehensive investigation. Here, we resolve distinct orders of Brown-Zak oscillations~\cite{Brown1964,Zak1964a,Zak1964b} whose periodicities are independent of the carrier density $n_c$.\\

This phenomenon is fundamentally analogous to the formation of the Hofstadter butterfly fractal spectrum in moiré superlattices~\cite{Yankowitz2012,Cao2018,Kim2017}. In the presence of both a magnetic field and a periodic arrangement of scatterers, the spectral gaps are described by the Wannier Diophantine equation~\cite{Dean2013}:
\begin{equation}
(n_c/n_a)=t\,\beta+s\ ,
\label{eq:diophantine}
\end{equation}
where $\beta = \phi/\phi_0$ is the ratio between the magnetic flux $\phi$ per unit-cell area and the magnetic flux quantum $\phi_0 = h/e$, $t$ and $s$ are integer Chern and Bloch band filling factor indices, and $n_a$ represents the nominal lithographic density of antidots in a given patterned region. Rearranging Eq.~\eqref{eq:diophantine} yields:
\begin{equation}
B=\frac{\phi_0}{t}\left(n_c-s\,n_a\right)\ .
\label{eq:B_general}
\end{equation}
At fixed carrier density and for the primary spectral gap ($t=1$), adjacent Wannier replicas ($s\rightarrow s+1$) yield a constant magnetic-field interval within consecutive Brown-Zak peaks:
\begin{equation}
\Delta B = B(s+1)-B(s) = \phi_0 n_a\ .
\label{eq:deltaB_general}
\end{equation}
Physically, $\Delta B$ represents the magnetic-field increment required to introduce exactly one additional flux quantum per unit-cell area.\\

Crucially, in the antidot superlattices investigated here, $n_a$ must be replaced by a dynamic, field-dependent effective density $n_k$. As schematically illustrated in Fig.~\ref{fig:Zak}, changing the magnetic field systematically tunes the semiclassical cyclotron radius $R_\mathrm{C} \propto 1/B$, forcing carrier trajectories to lock into distinct commensurability configurations $\mathrm{H}k$. Each orbit family encloses a cluster of scatterers, defining an effective superlattice unit cell of area $S_k = k a^2$ (where $k \in \{1, 2, 4, 9\}$). Consequently, the effective scatterer density perceived by the carriers is not merely a static lithographic parameter of the nanopatterned region, but rather a dynamic quantity $n_k = 1/S_k = n_a/k$ dictated by the active magnetic field scale. The underlying periodic potential experienced by Bloch electrons effectively transforms as $B$ varies, altering the effective Bravais lattice.\\

Driven by this magnetic-field-dependent restructuring of the probed unit cell, the magnetic flux enclosed by each effective superlattice region scales proportionately, mapping an orbital hierarchy of $\beta_{{\rm H} k} = k\beta_{\mathrm{H1}}$. Thus, Brown-Zak oscillations superimposed on each Weiss peak undergo a discrete, fractional rescaling:
\begin{equation}
\Delta B_{{\rm H}k} = \phi_0 n_k = \frac{\phi_0}{S_k} = \frac{\Delta B_{\rm H1}}{k}\ ,
\label{eq:rescaled_deltaB}
\end{equation}
corresponding directly to the effective unit-cell areas $a^2$, $2a^2$ (rotated by $45^\circ$), $4a^2$, and $9a^2$ experienced by the carriers on top of the dominant commensurability orbits $\mathrm{H1}$, $\mathrm{H2}$, $\mathrm{H4}$, and $\mathrm{H9}$ (Fig.~\ref{fig:Zak}).\\

\begin{figure*}[htbp]
    \centering
    \includegraphics[width=\textwidth]{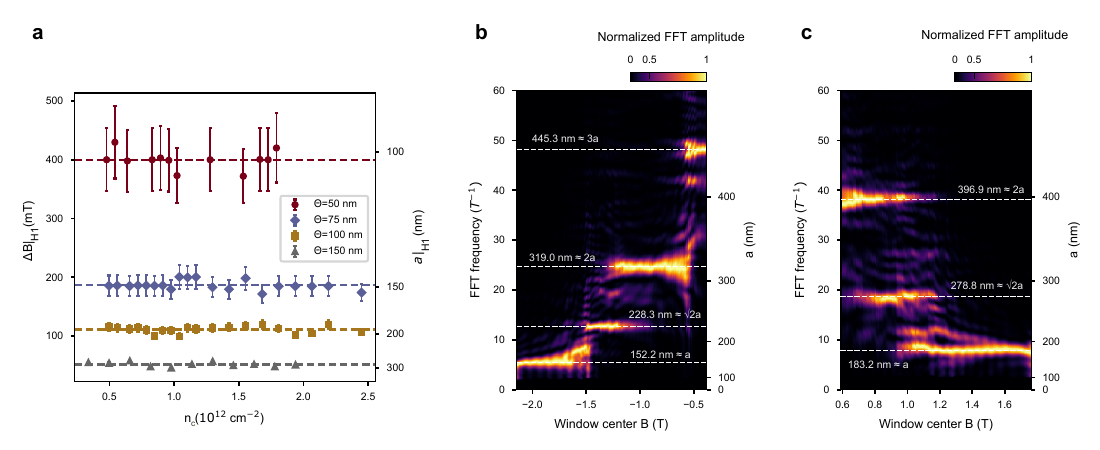}
    \caption{\textbf{Sliding-window Fourier analysis and structural scaling of superlattice oscillations.} \textbf{a)}~Experimental oscillation spacing $\Delta B$ at the $\mathrm{H1}$ peak as a function of carrier density for four different antidot dimensions ($\Theta = 50, 75, 100, 150\,\rm nm$ from Sample A). The density-independent behavior confirms the structural nature of the oscillations. Horizontal dashed lines indicate the mean spacing $\overline{\Delta B_{\rm H1}}$ used to extract the effective superlattice parameters $a_{|\rm H1}$ indicated on the right-hand axis. \textbf{b)}~Sliding-window FFT analysis of the longitudinal magnetoresistance for $\Theta = 75\,\rm nm$ versus the magnetic field window center. The right-hand axis shows extracted real-space periodicities matching $a$, $\sqrt{2}a$, $2a$, and $3a$. Prior to the FFT, the classical Weiss background (Fig.~\ref{normalizedcomm} inset) was subtracted, and individual FFT columns were normalized to their peak amplitude. \textbf{c)}~Sliding-window FFT spectrum for $\Theta = 100\,\rm nm$, resolving stable frequency bands at $a$, $\sqrt{2}a$, and $2a$.}
    \label{fig:FFTs}
\end{figure*}

%To systematically analyze the period of the fine-structure ripples observed on top of the main magnetoresistance peaks ($\Delta B$), we evaluate the Fast Fourier Transform (FFT) within a sliding magnetic field window of width $B_{\mathrm{win}}$ (Fig.~\ref{fig:FFTs}). This analysis yields the oscillation frequency $f = 1/\Delta B$ with a window-limited frequency resolution $\Delta f = 1/B_{\mathrm{win}}$, corresponding to a period uncertainty of $\Delta B^2/B_{\mathrm{win}}$ (see error bars).\\

To systematically analyze the period of the fine-structure ripples observed on top of the main magnetoresistance peaks ($\Delta B$), we evaluate the Fast Fourier Transform (FFT) within a sliding magnetic field window (details on the procedure are found in Methods). Figure~\ref{fig:FFTs}(a) focuses on the primary $\mathrm{H1}$ peak, presenting the average ripple spacing $\Delta B$ across different patterned regions over a carrier density range of $n_c = (0.50 - 2.25) \times 10^{12}\,\rm cm^{-2}$. The peak spacing remains constant across all $n_c$, confirming that the periodicity is an intrinsic property of the spatial array rather than an electronic effect. Horizontal dashed lines indicate the mean spacing $\overline{\Delta B}$ for each region. Substituting $\overline{\Delta B_{\rm H1}}$ into Eq.~\eqref{eq:rescaled_deltaB} for $k=1$ ($n_1 = n_a$) yields fundamental lattice parameters of $a|_{\mathrm{H1}} = 101.7\,\rm nm$ ($\Theta = 50\,\rm nm$), $149.0\,\rm nm$ ($\Theta = 75\,\rm nm$), $192.8\,\rm nm$ ($\Theta = 100\,\rm nm$), and $282.7\,\rm nm$ ($\Theta = 150\,\rm nm$), displaying excellent quantitative agreement with the lithographic dimensions across all patterned regions.\\

Beyond the $\mathrm{H1}$ magnetoresistance peak, Figs.~\ref{fig:FFTs}(b) and (c) display Fourier maps with flat, highly stable frequency bands corresponding to larger real-space periodicities: $a$, $\sqrt{2}a$, $2a$, and $3a$ (white dashed lines). These discrete frequencies reflect the reduction of $n_k$ across higher-order Weiss windows, exactly matching Eq.~\eqref{eq:rescaled_deltaB} and Fig.~\ref{fig:Zak}. This multi-tiered response originates from the quantum mechanical symmetry restoration governed by the Magnetic Translation Group (MTG), formulated in the seminal works of Brown and Zak~\cite{Brown1964,Zak1964a,Zak1964b}. In the presence of a magnetic field, the standard spatial translation operators of a crystal lattice no longer commute because electrons accumulate an Aharonov-Bohm phase along closed paths. Instead, the magnetic translation operators $\widehat{T}_M(\mathbf{R})$ satisfy a non-commutative ray group relation:
\begin{equation}
\widehat{T}_M(\mathbf{R}_1)\widehat{T}_M(\mathbf{R}_2) = e^{i 2\pi \beta(\mathbf{R}_1, \mathbf{R}_2)}\widehat{T}_M(\mathbf{R}_2)\widehat{T}_M(\mathbf{R}_1)\ ,
\end{equation}
where $\mathbf{R}_1$ and $\mathbf{R}_2$ represent two non-collinear basis vectors of the lattice that span an effective unit cell, and $\beta(\mathbf{R}_1, \mathbf{R}_2)$ is the precise flux ratio enclosed by that parallelogram. Zak demonstrated that commutativity is restored, and the irreducible representations of the MTG remain finite-dimensional, only when this enclosed flux forms a rational ratio with the flux quantum $\phi_0$ (i.e., when $\beta$ is a rational integer or fraction). When this rationality condition is fulfilled, the MTG becomes isomorphic to the ordinary commuting translation group, periodically restoring a delocalized, Bloch-like character to the charge carriers. In our system, as $B$ shifts the dominant trajectory to enclose larger effective unit cells $S_k$, the effective density $n_k$ steps down discretely, giving rise to the rescaled magnetic field quantization conditions expressed in Eq.~\eqref{eq:rescaled_deltaB}.\\

While flux quantization dictates the structural Brown-Zak periodicities through $n_k$, the broad Weiss commensurability peaks act as dynamical visibility windows. Transport within each window is dominated by trajectories sampling a specific subset of the antidot lattice, selectively enhancing the Brown-Zak harmonic associated with that specific effective density $n_k$. Semiclassical orbits thus do not set the Brown-Zak period through their cyclotron area; rather, they modulate the visibility of density-independent structural harmonics by selecting which effective superlattice unit cell $S_k$ dominates carrier transport.\\

\begin{figure*}
    \centering
     \includegraphics[width=\linewidth]{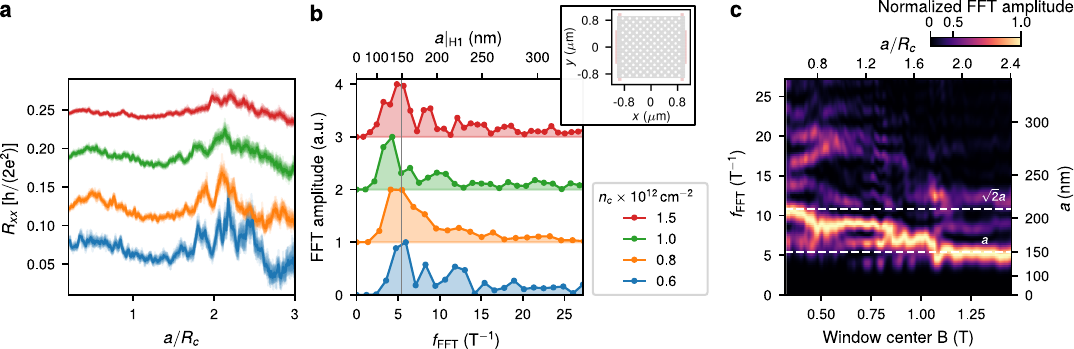}
    \caption{\textbf{Numerical quantum transport results for a matching antidot lattice ($a = 150\,\rm nm$).} \textbf{a)}~Longitudinal resistance $R_{xx}$ versus inverse cyclotron radius ($a/R_C$) for various carrier densities (vertically offset by $0.07\,h/2e^2$ for clarity). Thin lines denote 15 independent disorder realizations ($W = 0.025\,\rm eV$), while the bold trace represents the ensemble average. \textbf{b)}~Fourier transform of the ensemble-averaged $R_{xx}$ of the oscillation on top of the principal $\rm H1$ peak. Inset sketches the $9 \times 9$ antidot simulation layout. 
    \textbf{c)}~Sliding-window FFT for $n_c = 0.6\times10^{12}\,\mathrm{cm^{-2}}$, showing a decreasing of the effective lattice when lowering the magnetic field.}
    \label{fig:theoryRxx}
\end{figure*}

To validate this mechanism in a fully quantum-interference formalism, we performed tight-binding transport calculations within the Landauer-B\"uttiker formalism for the holey graphene Hall bar sketched in the inset of Fig.~\ref{fig:theoryRxx}(b). Weak Anderson disorder is included to mimic experimental imperfections; see Methods for numerical details.
Figure~\ref{fig:theoryRxx}(a) shows the calculated magnetoresistance $R_{xx}$ versus the normalized field $a/R_\mathrm{C}$ for a range of carrier densities, reproducing the features observed experimentally: Weiss commensurability peaks with superimposed \textit{wrinkles}. Panel (b) shows the Fourier transform of these superimposed oscillations within the main commensurability peak $\mathrm{H1}$. For all densities the spectrum is dominated by a peak corresponding to $a|_{\mathrm{H1}} \approx a$, matching the real-space
periodicity of the antidots, and its position is largely independent of $n_c$, confirming the structural origin of the \textit{wrinkles}.
Panel (c) shows a sliding-window Fourier analysis for the lowest carrier density, $n_c = 0.6\times10^{12}\,\mathrm{cm^{-2}}$, where the oscillations are most pronounced. The dominant frequency changes with field: within the $\mathrm{H1}$ window ($a/R_\mathrm{C} \gtrsim 1.8$) it settles at $a|_{\mathrm{H1}} \approx a$,
whereas at lower fields it shifts to a larger effective period. This field-dependent change of the dominant periodicity, tied to the commensurability window being probed, qualitatively reproduces the rescaling observed experimentally.\\

The observation of these fractional Brown-Zak components confirms that the artificial periodic potential is sufficiently regular for carriers to probe the translational symmetry of the effective superlattices $S_k$ over multiple unit cells, even in the presence of etched antidot boundaries.

\section{Summary}\label{Summary}

In this work, we have systematically investigated the magnetotransport properties of high-mobility encapsulated graphene antidot superlattices, examining the complex interplay between ballistic semiclassical trajectories, quantum interference, and artificial lattice dynamics. By scaling the lithographic antidot diameters across several dimensions and sweeping wide ranges of carrier densities and magnetic fields, we demonstrate a remarkably robust control over the electronic phase space.\\

Our transport measurements reveal multiple distinct families of commensurability (Weiss) peaks arising from the coexistence of chaotic dynamics and regular pinned trajectories governed by Kac's lemma~\cite{Kac1947}.  Crucially, our results show that these commensurability features persist down to low carrier densities, challenging directly the established classical-to-quantum breakdown criterion $\lambda_F / 2\pi \ll D$. Because our devices remain firmly within the ballistic limit even at small feature sizes, we demonstrate that the transition between the quantum and classical transport regimes is significantly more gradual than previously reported, highlighting the structural integrity and minimal edge roughness of our etched arrays~\cite{Jessen2019}. \\

We resolve highly pronounced, density-independent fine-structure features superimposed directly onto the main commensurability peaks. Through a sliding-window FFT analysis, these features are unambiguously identified as Brown-Zak oscillations~\cite{Weiss1991, Barrier2020}. Our spectral analysis reveals flat, highly stable frequency bands that map directly onto the real-space geometric paths of a square Bravais lattice. This phenomenon is successfully modeled by a hierarchical self-similarity of nested, commensurate sub-lattices. As the cyclotron orbits expand macroscopically to encompass larger clusters of scatterers, the effective superlattice unit cells increase quadratically in area and the fundamental magnetic flux-quantization condition rescales discretely. 

The constant magnetic field spacing $\Delta B_{\rm{Hk}} = \phi_0/(ka^2)$ provides a direct, non-destructive probe to extract the effective lattice parameters, yielding exceptional quantitative agreement with our lithographic designs.\\

The exceptional resolution of these fractional flux sub-bands confirms that electronic phase coherence is strictly preserved across multiple artificial unit cells. Ultimately, these findings demonstrate that patterned graphene heterostructures provide an optimal, highly tunable platform for exploring Hofstadter-type fractal spectra and superlattice physics. By bypassing the strict sub-kelvin thermal demands or sub-nanometer alignment constraints inherent to moiré superlattices, this architecture offers a robust pathway for the tailored engineering of artificial electronic bands in 2D materials.\\

\section{Acknowledgments}\label{Acknowledgments}

We wish to acknowledge scientific discussions with A. Hamilton, R.A. Jalabert and D. Weinmann.

\section{Funding}\label{Founding}
This work was supported by the European Union, Agencia Estatal de Investigaci\'{o}n of Spain (Grant PID2022-136285NB-C31/C32) and FEDER/Junta de Castilla y León Research (Grant SA106P23). P. C. and J.~E.-A. acknowledge support from the Spanish Ministerio de Ciencia, Innovaci\'{o}n y Universidades (Grants FPU24/01634 and FPU22/01039). A.P.R.
acknowledges the financial support received from
the Marie Skłodowska Curie-COFUND program
under the Horizon 2020 research and innovation initiative of the European Union, within the
framework of the USAL4Excellence program (Grant
101034371).   K.W. and T.T. acknowledge support from the CREST (JPMJCR24A5), JST and World Premier International Research Center Initiative (WPI), MEXT, Japan.

\section{Author Contribution}

J.S.-S., C.S.-S. and V.C. fabricated the samples. P.C. characterized all devices with help from M.A., J.S.-S., A.P.-R. and E. Diez. All results were analyzed and interpreted by P.C., with help from J.E.-A., Y.B., J.S.-S, and M.A. and inputs from the rest of authors. Y.B. perfomed the numerical the simulations with help from R.A.M.\textit{h}-BN samples were provided by K.W. and T.T. M.A. and P.C. wrote the manuscript with input from all authors. F.D.-A., E.Diez., E.D\'{i}az and M.A. conceived and supervised the project.

\bibliography{bibliography}

\newpage

\section{Methods}

\subsection*{Device fabrication} \label{sec:DeviceFab}
Heterostructures were fabricated via standard mechanical exfoliation of flakes from pristine crystals \textit{h}-BN and graphite. For Sample A the top and bottom \textit{h}-BN layers had thicknesses of $45\, \rm nm$ and $12.5\, \rm nm$, respectively, while for Sample B they were $60$ nm and $8$ nm. Thicknesses were measured using a Bruker Nano DektakXT profilometer, and monolayer graphene was confirmed through micro-Raman spectroscopy.
Heterostructure assembly was performed using a polycarbonate (PC) film supported on polydimethylsiloxane (PDMS). The top \textit{h}-BN was picked up at $50$–$60^\circ$C and transferred onto the graphene monolayer at $190^\circ$C. The stack was later placed on the bottom \textit{h}-BN, which was previously annealed in vacuum at 350$^\circ$C to clear any remaining residues that could worsen the contact surface. A full micro-Raman map of the completed heterostructure was used to identify clean and defect-free regions for patterning. Following heterostructure fabrication, a premask process was carried out using EBL to clean the surrounding area and avoid shorts between electrodes. A homemade PMMA resist (5\% in chlorobenzene for the premask and Hall bar) was spin-coated onto the sample, and a standard SF$_6$ etching process at room temperature was used to selectively remove the undesired material.\\

Once the premask process was done, the heterostructures were shaped into Hall bars using EBL, with Sample A forming a 16-terminal layout [see Fig.~\ref{superfigure1}(a)] and Sample B, a 10-terminal one. A final round of EBL and cryo-etching~\cite{Clerico2019,EBLVito2020} was used to define periodic antidot arrays within the Hall bars, using here a 2\% in chlorobenzene resist, thinner than the one used for the premask and Hall bar, to ensure a better definition of the smooth edges that support nearly specular electron reflection~\cite{Estrada2025}. Cryo-etching, which combines low-temperature plasma etching and physical sputtering, proved essential for achieving a clean, sharp definition of the antidot superlattices, ensuring the formation of a clear periodic potential necessary for miniband development. To optimize antidot definition, dose array calibrations were performed on sacrificial \textit{h}-BN flakes, replicating the final design [see SEM micrographs shown in Fig.~\ref{superfigure1}(a)]. Doses ranging from $200$ to $600\,\mu$C/cm$^2$ were tested, and $350$, $375$, $400$ and $425\,\mu$C/cm$^2$ yielded optimal resolution for $50$, $75$, $100$ and $150\,$nm antidots, respectively. Metal contacts ($10/55\,$nm Cr/Au) were then deposited by e-beam evaporation and subsequently bonded on an LCC20 chip carrier for electrical characterization.

\subsection*{Electrical characterization} \label{sec:ElectricalChar}
All electrical measurements were carried out in an Oxford Instruments Teslatron cryostat equipped with a 360$^\circ$ mechanical rotator, providing a base temperature of $300\,$mK and magnetic fields up to $12\,$T. Unless otherwise stated, measurements were performed at 1.5 K. Standard low-frequency lock-in techniques were employed: a reference AC current of $50\,$nA was injected through a $100\,$M$\Omega$ series resistor at a frequency of $11\,$Hz. The back-gate voltage was controlled using a Keithley 2612 sourcemeter, enabling stable and precise tuning of the carrier density during all measurements.

\subsection*{Brown-Zak oscillations analysis}\label{sec:FFT}

To systematically analyze the period of the fine-structure ripples observed on top of the main magnetoresistance peaks ($\Delta B$), we evaluate the FFT of the oscillations obtained by subtracting a smooth curve corresponding to the classical contribution (Weiss oscillation) within a sliding magnetic field window of width $B_{\mathrm{win}}$. This analysis yields the oscillation frequency $f = 1/\Delta B$ with a window-limited frequency resolution $\Delta f = 1/B_{\mathrm{win}}$, corresponding to a period uncertainty of $\Delta B^2/B_{\mathrm{win}}$ (see error bars Fig.~\ref{fig:FFTs}).

\subsection*{Details on the tight-binding calculations} \label{sec:AppTheory}
For the numerical calculations, we employ a scaled tight-binding Hamiltonian for graphene~\cite{Liu2015, Mrenca2023}. 
The scaled graphene model is obtained by considering an enlarged carbon-carbon bond length $a = a_0 s_f$, with $a_0 = 0.142 \si{nm}$ being the real graphene value and $s_f>1$ the scaling factor. The tight-biding is then built in an enlarged hexagonal lattice with lattice constant $a_{\rm hex} = \sqrt{3}a$ and rescaled hoping parameter $t=t_0/s_f$, where $t_0=2.8\si{eV}$ is the graphene hoping parameter of pristine graphene and $s_f$ ensures that the Fermi velocity of the Dirac cones of the effective model $\hbar v_F = (3/2) a t$ is the same as in the non-scaled case, namely $(3/2)a_0t_0$. Using the scaled model, it is possible to construct systems on a scale similar to that of the experiments, considerably reducing the required numerical effort. 
The tight-binding Hamiltonian is given by
\begin{equation} \label{AppTh:eq:Ham}
\mathcal{H} = 
    \sum_i (\mu + U_{i})c^\dagger_i c_i^{}
    - \frac{t_0}{s_f}\sum_{\langle i,j \rangle }   e^{i\varphi_{i,j}}c^\dagger_i c_j^{}\ ,
\end{equation}
where $c^\dagger_i$ ($c_i^{}$) creates (destroys) an electron at site $i$ and $\langle i,j \rangle $ indicates a sum over first nearest neighbors. In the previous expression $\mu$ is the chemical potential, $U_i \equiv U(x_i, y_i)$ is an on-site space-dependent potential. The Peierls phase $\varphi_{i,j}$ is included to take into account the external magnetic field $\mathbf{B}$ and it is given by $\varphi_{i,j} = - (2 \pi /\phi_0) \int_{\mathbf{r}_i}^{\mathbf{r}_j} d \mathbf{r} \cdot \mathbf{A}(\mathbf{r})$  where $\mathbf{A}$ is the vector potential such that $\mathbf{B} = \nabla\times \mathbf{A}$. 

We set the vector potential in the Landau gauge as $\mathbf{A} = (B_z y, 0, 0)$ such that enters only in the hoping terms in the $x$ direction. 
The leads are defined as infinite graphene nanoribbons defined from the periodic-boundary version in $x$ or $y$-direction of the Hamiltonian~\eqref{AppTh:eq:Ham} with $U_i \to 0$ and $\mu\to \mu_{\rm lead}$. 
To prevent unwanted scattering at the interfaces with the vertical leads, a phase shift is included $\varphi_{i,j} = -2\pi/\phi_0 B_z y_{\rm lead}(x_j-x_i)$, where $y_{\rm lead}$ is the coordinate of where the lead intersects the scattering region. 

In the simulations we include uncorrelated disorder by an Anderson term $U_i = (W/2) u_i$, where $W$ is the strength of the disorder and $u_i \in [-1,1)$ is a random number from a uniform distribution, different for each site and disorder realization. In Fig. \ref{fig:theoryRxx} the magnitude of disorder is set to $W = 0.025 \si{eV}$. Qualitatively similar results were obtained by changing this value as along as $W/\mu \ll 1$. 

The scattering problem is solved employing the \texttt{Kwant} toolkit~\cite{Groth2014} by obtaining the transmission probability $T_{ij}$ from the incoming modes in leads $j$ to the outgoing modes in leads $i$. The transmission probabilities are related to the conductance through the Landauer–Büttiker formula
\begin{equation} \label{AppTh:eq:Landauer}
 I_p = \sum_q G_{pq} [V_p - V_q]  ~,
\end{equation}
with $G_{pq} = (2e^2/h) T_{pq}$. We solve the system of equations by inverting it and setting the current between two fixed leads and solving the voltages $V_p$  for all leads. 

Then, the resistance between voltage probes $i$ and $j$ for a current between leads $k$ and $l$ is defined as: 
\begin{equation}
    R_{ij; kl} = \frac{V_i - V_j}{I_k-I_l}~. 
\end{equation}
In particular, the longitudinal resistance is defined in our geometry by setting the current between the two horizontal leads $I_{\rm left} = - I_{\rm right}$ and by considering the voltage drop between the two consecutive vertical leads at the upper end of the scattering region. We have checked that a consistent results is obtained by reversing the current and considering the voltage drop at the two vertical leads located in the lower end of the setup.

\par The calculations are done at fixed chemical potential in the leads $\mu_{\rm lead} = 0.12 \si{eV}$ and by changing the chemical potential of the scattering region to match the carrier density according to $\mu = \hbar v_F \sqrt{\pi n_c}$. In particular, the values reported in Fig.~\ref{fig:theoryRxx} correspond to $\mu \approx [0.08, 0.09, 0.11, 0.13] \si{eV}$. 

\end{document}